\documentstyle[11pt,pasp,twoside]{article}
\def\edcomment#1{\iffalse\marginpar{\raggedright\sl#1\/}\else\relax\fi}
\reversemarginpar

\begin{document}
\title{The peculiar X-ray properties of the low-redshift quasar MR\,2251-178 
                                  and its environment: some surprises}
 \author{Stefanie Komossa}
\affil{Max-Planck-Institut f\"ur extraterrestrische Physik, Giessenbachstr. 1, D-85748 Garching,
       Germany; skomossa@mpe.mpg.de}

\begin{abstract}
MR\,2251-178 was the first quasar initially discovered in X-rays,
and the first one found to host a warm absorber.
Despite its many outstanding multi-wavelength properties,
MR\,2251-178 has not yet been studied in detail with
recent X-ray observatories.
Here, we present results from an
analysis of the X-ray spectral, temporal, and spatial
properties of this source and its environment
based on deep observations carried out with the X-ray satellite {\sl ROSAT}.
The derived mean {\em soft} X-ray luminosity of MR\,2251-178, $\sim10^{45}$ erg/s,
places the quasar among the most X-ray luminous AGN in the local universe.   
\end{abstract}

\section{Introduction}

MR 2251-178 at redshift $z$=0.064 was 
the first quasar initially discovered by X-ray observations
(Ricker et al. 1978),
and the first one found to host an ionized absorber (Halpern 1984).
The quasar turned out to be outstanding in many respects.
It has a high ratio of $L_{\rm x}/L_{\rm opt}$, is
surrounded by the largest quasar emission line nebula known,
and is located in the outskirts of a cluster of galaxies  (e.g., Bergeron
et al. 1983). Our analysis of the {\sl ROSAT} PSPC observations of this source
and its environment revealed a number of surprises, summarized below.

\section{X-ray results: some surprises}
Results presented here are based on the analysis of two {\sl ROSAT} PSPC
observations of MR\,2251-178, carried out in Nov. 1990 (all-sky survey)
and Nov. 1993 (18.3 ksec duration). Details of the data analysis 
and the discussion of results is presented by Komossa (2001).  
The following results were obtained:

\paragraph{X-ray variability.} There is evidence for an X-ray flaring event with
variability by a factor $\sim$2 within 10 ksec during the {\sl ROSAT} all-sky survey;
rarely observed in such a luminous source, but similar to PDS\,456 (Reeves et al. 2000)
and PKS\,0558-504 (Wang et al. 2001).

The source was a factor $\sim$3 brighter during the later pointed
{\sl ROSAT} observation. The mean observed soft X-ray luminosity
at that epoch, $L_{\rm (0.1-2.4)keV} \simeq 10^{45}$ erg/s,
places the quasar among the most X-ray luminous AGN in the local
universe.

\paragraph{X-ray spectrum.} Remarkably, we do not detect any excess X-ray {\em cold} absorption
expected to originate from the giant [OIII] gas nebula surrounding MR\,2251-178.
This finding excludes the presence of a huge HI extent of the HII
emission line gas (along the line-of-sight),
and constrains some formation scenarios
(e.g., Shopbell et al. 2000) of the gas nebula.

On the other hand, we do find that the X-ray spectrum is 
modified by lots of amounts of gas along the 
line of sight - but this material is {\em highly ionized}. 
As indicated by some, but not all,
earlier X-ray observations of MR2251-178,
a single powerlaw does not provide an acceptable
spectral fit ($\chi^2_{\rm red}=5.4$).
The presence of an {\em ionized absorber}  markedly improves
the fit. We find an ionization parameter
$\log U = 0.5$ and a column density
$\log N_{\rm w} = 22.6$ of the ionized material
(see Komossa 1999 for a general review on warm absorbers).
The presence of highly ionized material along the
line of sight is consistent with the detection of 
absorption lines in the UV spectrum of MR\,2251-178
(Monier et al. 2001). 
Our best-fit warm-absorber model still leaves some residuals
at the low-energy part of the spectrum, suggesting the presence
of a very soft excess.

\paragraph{Surrounding sources.} None of the other 
optically bright member galaxies of the cluster
to which MR\,2251-178 belongs, are detected in X-rays.
However, east of the quasar there is a significant excess of
X-ray sources (a factor of 4 compared to the log N -- log S
distribution of Hasinger et al. 1994), 
several of them
without optical counterparts on the UK Schmidt plates.

\paragraph{Intra-cluster gas emission:} The X-ray emission from the intra-cluster medium is
weak or absent. We derive an upper limit on the X-ray luminosity
of $L_{\rm x} \le 2\,10^{42}$ erg/s from the direction
of the optical center of the cluster,  weaker than other clusters
of comparable richness. This may indicate that what appears
as one single cluster could be just a chance projection
of several poorer clusters or groups. Alternatively, the cluster
X-ray emission might be off-set from the optical center.


\begin{references}

\reference Bergeron J., et al. 1983, MNRAS 202, 125

\reference Halpern J. 1984, ApJ 281, 90

\reference Hasinger G., et al. 1994, A\&A 275, 1

\reference Komossa S. 1999, in {\sl ASCA-ROSAT workshop on AGN}, ISAS Report,
                T. Takahashi \& H. Inoue (eds), p. 149 [also available at astro-ph/0001263]


\reference Komossa S. 2001, A\&A 367, 801 

\reference Monier E.M., et al. 2001, ApJ, submitted 

\reference Reeves J.N., et al. 2000, MNRAS 312, L17  

\reference Ricker G.R., et al. 1978, Nature 271, 35 

\reference Shopbell P.L., Veilleux S., Bland-Hawthorn J. 1999, ApJ 524, L83

\reference Wang T., et al. 2001, ApJ, in press 

\end{references}
\end{document}